# Propagation of Visible Light in Nanostructured Niobium Stripes Embedded in a Dielectric Polymer

F. Telesio[1], F. Mezzadri[2,3], M. Serrano-Ruiz[4], M. Peruzzini[4], F. Bisio[5], S. Heun[1], and F. Fabbri[1]

*1 NEST, Istituto Nanoscienze – CNR and Scuola Normale Superiore, Piazza San Silvestro 12, 56127 Pisa, Italy*

*2 IMEM-CNR, Parco Area delle Scienze 37/a, Parma, 43124, Italy*

*3 Department of Chemistry, Life Sciences and Environmental Sustainability, University of Parma, Parco Area delle Scienze 11/A, 43124 Parma, Italy*

*4 CNR-ICCOM, Via Madonna del Piano 10, 50019 Sesto Fiorentino, Italy*

*5 CNR-SPIN, C.so Perrone 24, 16152 Genova, Italy*

Corresponding author: <u>francesca.telesio@nano.cnr.it</u>, <u>filippo.fabbri@nano.cnr.it</u>

**Abstract**

Nanometric metallic stripes allow the transmission of optical signals via the excitation and propagation of surface-localized evanescent electromagnetic waves, with important applications in the field of nano-photonics. Whereas this kind of plasmonic phenomena typically exploits noble metals, like Ag or Au, other materials can exhibit viable light-transport efficiency. In this work, we demonstrate the transport of visible light in nanometric niobium stripes coupled with a dielectric polymeric layer, exploiting the remotely-excited/detected Raman signal of black phosphorus (bP) as the probe. The light-transport mechanism is ascribed to the generation of surface plasmon polaritons at the Nb/polymer interface. The propagation length is limited due to the lossy nature of niobium in the optical range, but this material may allow the exploitation of specific functionalities that are absent in noble-metal counterparts.

**Introduction**

Light can be transported in metal stripes or wires thanks to the excitation and propagation of surface plasmon polaritons (SPP), resonant oscillations of the free-electron plasma at the interface between a metal and a dielectric coupled to a light field [1–3]. A SPP is a surface electromagnetic (EM) wave, whose associated field is confined to the near vicinity of the dielectric–metal interface. This confinement leads to an enhancement of the EM field at the interface, resulting in an extraordinary sensitivity of SPPs to surface conditions. SPP-based devices exploiting this sensitivity are thus widely used as chemo- and bio-sensors [4,5]. The enhancement of the EM field at the interface is responsible for surface-enhanced optical phenomena such as Raman scattering (SERS), second harmonic generation (SHG), fluorescence, etc. [6,7]. The intrinsically two-dimensional nature of SPPs provides significant flexibility in engineering SPP-based all-optical integrated circuits needed for optical communications and computing [8–11]. The relative ease of manipulating SPPs on a surface opens an opportunity for their application in photonics and optoelectronics [12].

Despite this wealth of appealing applications, SPP based waveguides suffer from high SPP damping due to Ohmic losses that limit the propagation length to the order of tens (in the visible range) or hundreds (in the near-infrared) of micrometers. Recently, an innovative approach in the fabrication of SPP-based waveguides has replaced inorganic dielectrics (such as $SiO_2$ or $Si_3N_4$) with polymeric ones, in order to increase the propagation length in the range of technological interest, such as the telecom third window (1.55 µm) [13].

In order to limit the Ohmic losses, almost all existing plasmonic systems rely on noble metals, whose plasmonic properties are well-established [14]. Silver is one of the best candidates for nanoscale plasmonics, due to the lack of interband transitions in the visible range, low damping, and addressed plasma wavelength of 318 nm [15,16]. Most of the currenlty relevant devices rely on gold as the plasmonic material, which despite possessing relatively high optical damping in the visible due to strong interband transitions [15,17–19], is intrinsically inert, long-term stable, biocompatible, and easy to pattern. However, recent works have addressed the possible exploitation of lossy metals for developing SPP-based devices. Among them, niobium films are gaining increasing attention as an alternative plasmonic platform [20,21]. Niobium, which is considered a refractory plasmonic material, shows similar optical properties to Au in the near-infrared and mid-infrared[20]. In fact, niobium has proven suitable for high temperature plasmonic applications in the near and mid infrared, such as thermo-photovoltaics and thermal imaging.[22] In addition perfect plasmonic absorbers based on Nb nanostructures at cryogenic temperatures can be used as single photon detectors over a broad infrared spectral range.[21]

On the other hand, the use of niobium films is widespread in current high-sensitivity low temperature devices, by exploiting its superconducting transition (hence with negligible Ohmic losses), such as superconducting quantum interference devices (SQUIDs) and superconductor electronics [23,24]. Besides, niobium is thus a very versatile platform, and it is recently receiving increasing attention in the quantum technology community, due to its high critical field (Hc ~ 2 T) and critical temperature (Tc ~ 9.2 K).[25] High quality Josephson junctions based on Nb superconducting contacts were recently fabricated with different materials on the normal region, such as III-V semiconductors [26–30] and 2D materials [31,32], demonstrating good proximity effect and high transparency of interfaces.

In this work, we demonstrate the transport of visible light via the remote detection of the Raman signal of an exfoliated black phosphorus (bP) multilayer flake mediated by nanometric and nanocrystalline Nb stripes embedded in a dielectric polymeric layer. We ascribe the optical propagation to the generation of SPP at the interface between the metal and the dielectric. The Raman signal of black phosphorus is exploited as a probe for SPP-mediated light transport, allowing the assessment of this phenomenon without the need of separated light-injection and light-detection probes. Moreover, the sensitivity to air of bP flakes induced the need of coating, which is often made of PMMA [33,34], that introduce another interesting interface between the metal and a high-dielectric-constant medium. The accurate characterization of the morphological and structural properties of the niobium film reveals a nanostructured film composed of elongated grain. The propagation length is limited to a few micrometers by the lossy nature of the niobium in the optical range. However, a detailed comparison with a similar structure featuring gold stripes demonstrates superior light transport properties of the Nb stripes in the green-wavelength range (532 – 550 nm). Calculations based on optical constants retrieved from reference films confirm this seemingly counterintuitive results and suggest Nb as a functional alternative for plasmon-based light transport at short light wavelength.

**Experimental**

BP crystals were prepared following a well-established procedure [35]. The high reactivity with air of exfoliated bP represents a major challenge for the realization of high-quality devices. To overcome this issue, bP exfoliation was carried out in a glove bag under a nitrogen atmosphere. After the transfer of the flakes on a 300 nm $SiO_2$/Si wafer substrate, the samples were immediately coated with a protective layer of (poly(methyl methacrylate), PMMA), which is also the lithographic resist. The geometry of the devices was defined by electron beam lithography (EBL) using a Zeiss Ultraplus SEM equipped with Raith Elphy Multibeam software. The metallic stripes on the bP flakes are approximately 500 nm wide. After developing, a mild oxygen plasma of 10 W for 1 min with 40 sccm of oxygen was performed to efficiently remove all the resist residuals. Then the samples with Nb contacts were transferred to the vacuum chamber for the metal sputtering, and an in situ cleaning with Ar plasma was performed. The Ar cleaning conditions were tuned to have a gentle etching and a better control of the process. The etching time was calibrated to minimize the contact resistance. This last cleaning step was performed after the pre-sputtering of the two metallic targets for metal deposition, during which the sample was protected from contamination thanks to a closed shutter and a rotating carousel that allowed moving the sample away from the plasma source. Then 10 nm of Ti and 60 nm of Nb were deposited. After the sputtering process, the sample underwent a fast lift-off in acetone at 55 °C. Then it was immediately coated with a bilayer of copolymer methyl methacrylate (8.5) methacrylic acid (MMA (8.5) MAA) and PMMA, to guarantee protection from oxidation. This protection layer is approximately 700 nm thick, with 450 nm of copolymer and 250 nm of PMMA on top[32]. For the sample with Cr/Au stripes, after the oxygen plasma the sample was transferred into a thermal evaporator, and a 10 nm/65 nm Cr/Au bilayer was deposited. Then the sample underwent lift-off and all the following processes with the same timing and procedures as for the samples with Nb contacts. The superconducting properties of the Nb stripes produced in the aforementioned process are reported in previous works[26,30,32].

Scanning Raman spectroscopy was carried out with a Renishaw InVia system, equipped with a confocal microscope, a 532 nm excitation laser, and an 1800 line/mm grating (spectral resolution 2 $cm^{-1}$). All Raman data presented in this paper were obtained using the following parameters: excitation laser power 500 μW, acquisition time for each spectrum 15 s, pixel size 500nm x 500nm, 100X objective (NA=0.85). The laser spot size has been previously evaluated to be 800 nm in diameter [36]. A principal component algorithm is used to remove the noise. The Raman intensity maps are obtained by fitting the Raman modes with a Lorentzian peak considering a constant baseline.

The morphological analysis was carried out in Zeiss Merlin SEM, working at 15 kV of accelerating voltage and 150 pA of beam current and in Bruker Dimension Icon atomic force microscope (AFM), working in ScanAsyst mode.

Powder X-Ray diffraction data were acquired using a Rigaku Smartlab XE diffractometer. A CuK$_\alpha$ incident beam was parallelized through a parabolic mirror, and measurements were carried out in θ-θ geometry with 5° soller slits on both the incident and diffracted beam. A HyPix3000 detector was operated in 1D mode in the 10-110° 2θ range with 0.01° steps and 5°/min speed.

The dielectric function of Nb, from which a theoretical estimation of the SPP attenuation length was obtained, was measured on reference films by means of variable-angle spectroscopic ellipsometry (J.A. Woollam M2000, 245-1700 nm spectral range).

## Results

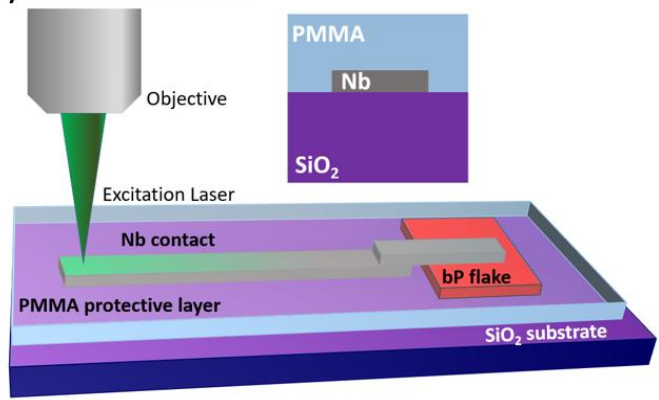
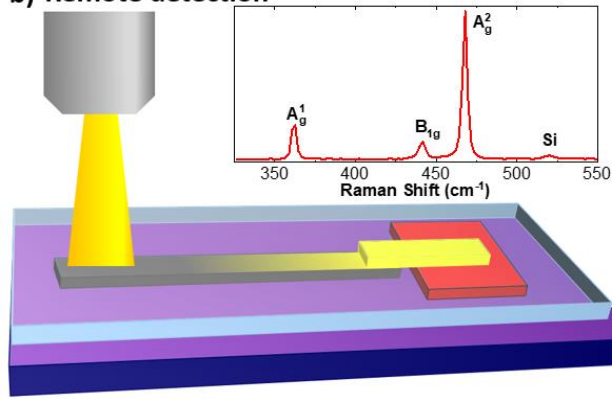

**Figure 1: Schematic representation of the experimental set-up for the optical excitation (a) and the Raman signal collection (b). The inset of the upper panel shows the cross-section of the structure under analysis, while the inset of the lower panel presents a representative Raman spectrum of the bP flake.**

Figure 1 presents sketches of the experimental set-up employed for the evaluation of light propagation in the niobium stripes. The experiment is set up as a standard Raman map, where the excitation laser is focused through the microscope objective while scanning the sample surface. When the laser is focused on the Nb stripe, we observed that it remotely excites the Raman phonon modes of bP. This demonstrates the light transport at the wavelength of the excitation laser (532 nm) (Fig. 1a). The remote detection of the Raman signal (Fig. 1b) is employed to assess the light transport at the wavelength of the Raman modes of bP (542 -546 nm). The remote excitation and remote detection of Raman modes have been previously reported in case of graphene interfaced with silver nanostructures. Nevertheless, this approach was not employed for the evaluation of the maximum propagation of light in the silver nanostructures [37]. This approach is novel for far-field optical transport measurements, albeit it is similar to the *scattered-light method* [38]. Indeed, it presents different advantages: first, it allows studying simultaneously the light propagation at two different wavelengths (excitation wavelength and Raman signal wavelength). Using a Raman scattering signal as probe allows to have a high intensity level, similar to the use of fluorescent molecules, without however the inconvenient of fluorescence molecular-bleaching [39]. The inset of Fig. 1a presents the cross section of the niobium stripes deposited on a 300 nm silicon dioxide substrate and capped with a multilayer of MMA (8.5) MAA copolymer and PMMA (thickness equal to 700 nm). A detailed characterization of the niobium thickness is reported in Fig. S1. The inset of Fig. 1b shows a representative Raman spectrum of a black phosphorus flake. Three peaks at 362 cm$^{-1}$ (542 nm), 441 cm$^{-1}$ (545 nm) and 467 cm$^{-1}$ (546 nm) correspond to the $A^1_g$, $B_{2g}$, and $A^2_g$ phonon modes of bP. They are attributed to the out-of-plane vibration ($A^1_g$) and the in-plane vibrations along the ZZ ($B_{2g}$) and AC ($A^2_g$) directions, respectively [40,41]. The faint peak at 520 cm$^{-1}$ is due to the silicon substrate.

In all experiments presented in this work, the laser polarization is kept perpendicular to the long edge of the bP flake, to account for the well-established angular anisotropy of the Raman bP signal [42–44].

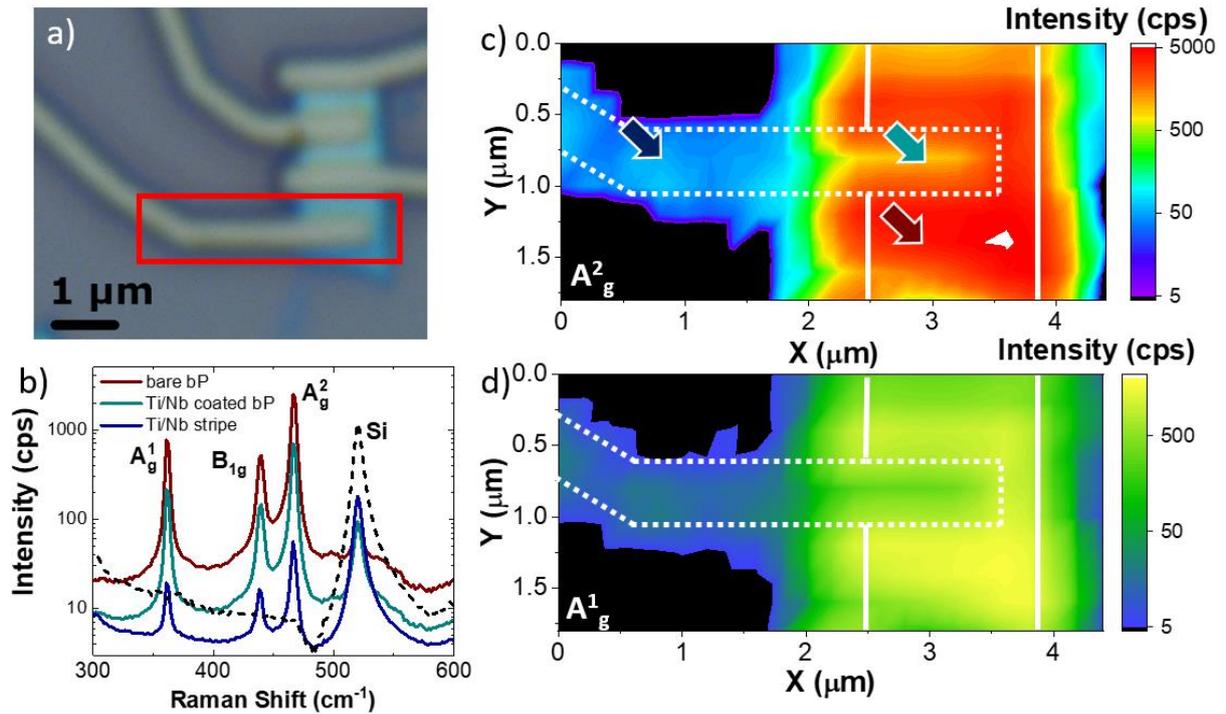

**Figure 2:** a) Optical microscopy image of the bP flake with Ti/Nb stripes. The red square indicates the area where the Raman maps were acquired. b) Raman spectra acquired on the bare bP (red line), the Ti/Nb coated bP (cyan line), and on the Ti/Nb stripe 1 μm away from the bP flake (blue line). The areas of collection are indicated in c) by the colored arrows. The same color code as for the spectra is employed. The spectrum of the SiO$_2$/Si substrate is shown as black dashed line for reference. The spectra are presented as acquired on a logarithmic scale with no shifting of the intensity. c) Intensity map of the $A^2_g$ mode. d) Intensity map of the $A^1_g$ mode.

Figure 2 presents the Raman data acquired according to the scheme reported in Fig. 1. Figure 2a shows the optical microscopy image of a bP flake with Ti/Nb stripes. The Raman spectra acquired on the bare bP (red line), the Ti/Nb-coated bP (cyan line), and on a Ti/Nb stripe (blue line), 1 μm away from the bP flake, are presented in Fig. 2b. The silicon peak is not discussed here, because silicon is present in every spectral acquisition.

The comparison between the red and cyan spectra reveals that the coating with Ti/Nb causes a decrease in the Raman modes intensity of the bP. The decrease in intensity is homogeneous for the different bP Raman modes. In fact, all Raman modes show an intensity decrease equal to 27%, probably due to the scattering and/or attenuation of the excitation laser. It is worth noting that acquiring a Raman spectrum on the Ti/Nb stripe 1 μm away from the bP flake, we are still able to detect the Raman modes of bP, demonstrating the optical transport of both the excitation probe and the Raman signal. The intensities of the Raman modes 1 μm away from the bP flake are 3% with respect to the bare bP intensities. In order to gain more insights, Raman intensity maps are acquired in the area highlighted by the red square in Fig 2a. The intensity maps of $A^1_g$ and $A^2_g$ peaks are shown in Fig 2c and 2d, respectively. Interestingly, we could clearly observe the fingerprint of the bP Raman spectrum also when the light is focused and collected along the Ti/Nb stripes, even outside the bP flake, whereas no bP signal can be observed outside the flake away from the Nb stripes. The Raman maps confirm that the intensity of bP Raman fingerprints are comparable along the Ti/Nb stripes. An additional analysis is carried out on Al/Nb stripes on a bP flake and reported in Fig. S2, while the morphological and structural properties of the Al/Nb films are reported in Fig S3. This analysis demonstrates the importance of the crystallinity of the Nb film, because the Al/Nb film demonstrates poor crystallinity, which causes a limited light propagation (less than 1 μm).

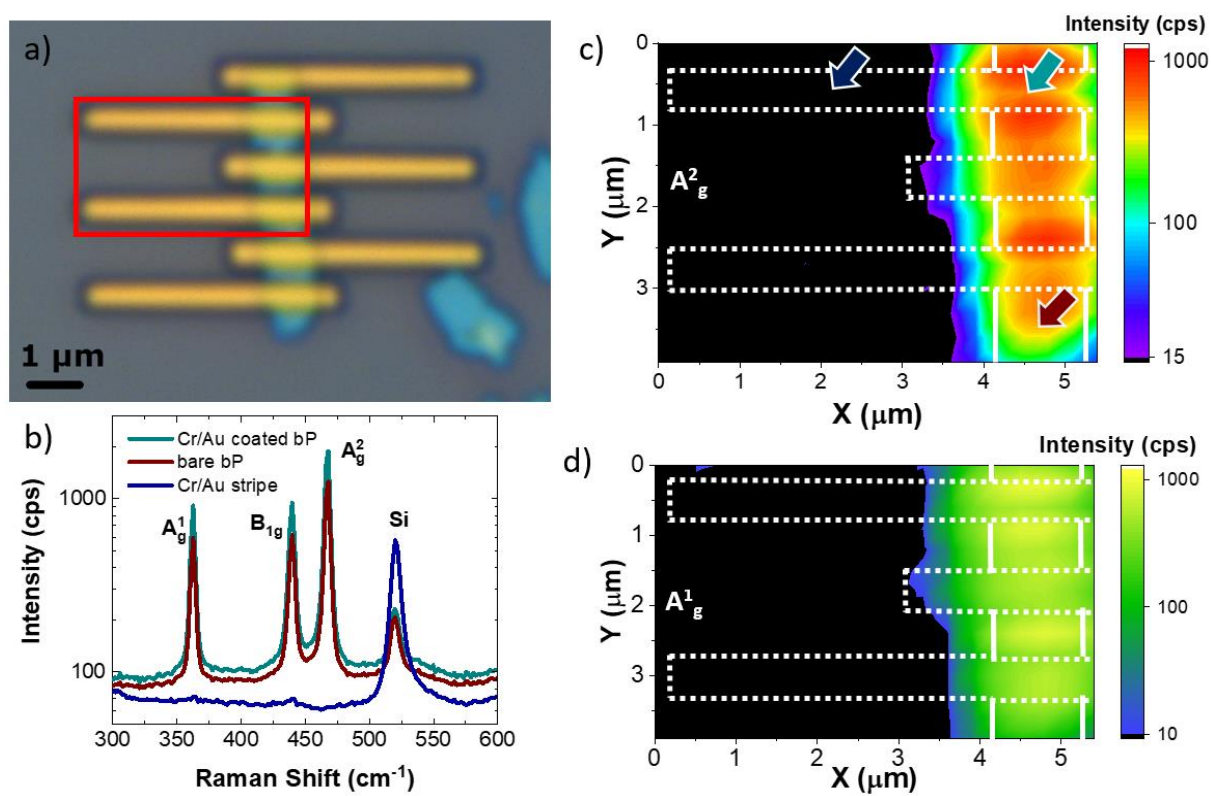

**Figure 3: a)** Optical image of a bP flake with Cr/Au stripes. The red square indicates the area where the Raman maps were acquired. **b)** Raman spectra acquired on the bare bP (red line), on the Cr/Au coated bP (cyan line),

and on the Cr/Au stripe at 1 μm away from the bP flake (blue line). The area of collection are highlighted by colored arrows in c) . The same color code as for the spectra is employed. c) Intensity map of the $A^2_g$ mode. d) Intensity map of the $A^1_g$ mode.

In order to rule out a possible light transport due to the geometry of the niobium stripes on $SiO_2$ capped by the dielectric polymer, we carried out similar experiments using gold stripes with an adhesion layer of chromium (Fig. 3). Figure 3a presents the optical microscopy image of the Cr/Au stripes deposited on the bP flake. The Raman spectra of the bare bP (red line), the Cr/Au coated bP (cyan line), and of the Cr/Au stripe, obtained 1 μm away from the bP flake, are presented in Fig 3b. All Raman modes of the bP present an equal increase of the intensity due to the coating with the Cr/Au. The intensity increase, equal to 66% of the intensity of the bare bP, is attributed to a surface enhanced Raman scattering (SERS) effect. The Raman spectrum (blue line) obtained 1 μm away from the bP flake does not show any Raman feature related to bP. The intensity maps of the $A^1_g$ and $A^2_g$ peaks, shown in Fig. 4c and 4d, respectively, show that in the case of Cr/Au stripes the bP Raman signal is not detected on the metal stripes away from the bP flake. Furthermore, the maps confirm the intensity increase of the Raman modes at the edge of the Au stripes coating the bP. Additionally, the smoother surface of the Au stripes in comparison with the Nb stripes promotes a weaker coupling between the SPP and the far-field radiation in free space.

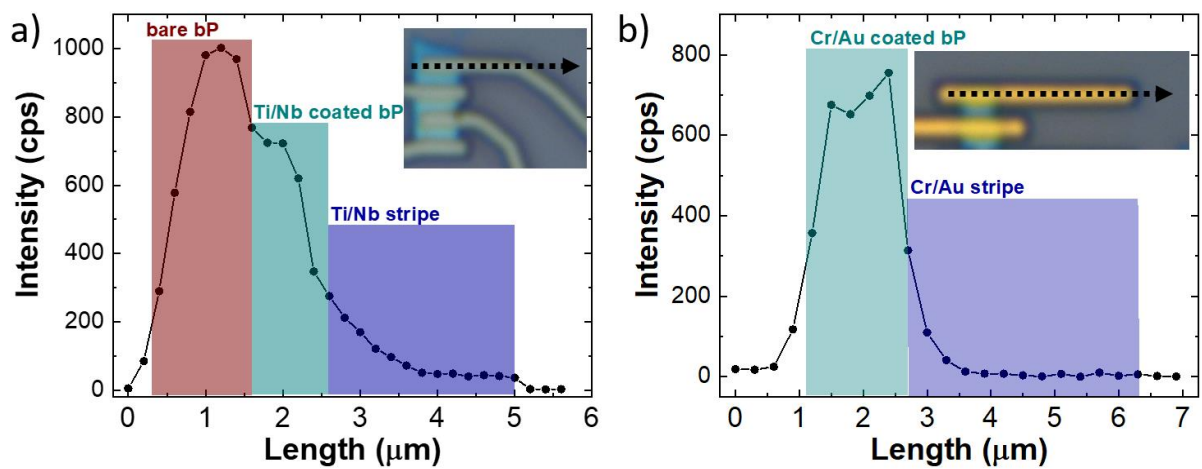

Figure 4: Evaluation of the attenuation length for (a) Ti/Nb and (b) Cr/Au stripes, using the $A^2_g$ mode intensity.

A careful evaluation of the detection length is obtained by using Raman line mapping along the Ti/Nb (Fig. 4a) and Cr/Au (Fig. 4b) stripes. The intensity of the $A^2_g$ Raman mode is employed as probe to evaluate the light detection length. In this particular case, we define the detection length as the maximum length at which we are still able to detect the Raman modes related to bP. In the case of the Ti/Nb stripes, the $A^2_g$ intensity first decreases due to the coating with the Ti/Nb (an intensity decrease comparable to that presented in Fig. 2), then it exponentially decays for 2.6 μm. We note that both the laser excitation and the Raman signal have to travel this distance. Therefore, the detection length is 2.6 μm at 546 nm ($A^2_g$ Raman mode wavelength), considering the convolution of the propagation of the excitation laser. In the case of the Cr/Au stripes, the Raman signal drops drastically to zero with a detection length that is equal to the exponential decay outside of the bP flake and equal to 600 nm. In order to have a more thoughtful evaluation of the light transport properties of the Ti/Nb and of the Cr/Au stripes, we obtained the attenuation length, using an exponential decay fitting of the data presented in Fig. 4 (see Fig. S4). The attenuation length is 600 nm and 280 nm in case of the Ti/Nb and Cr/Au stripe, respectively.

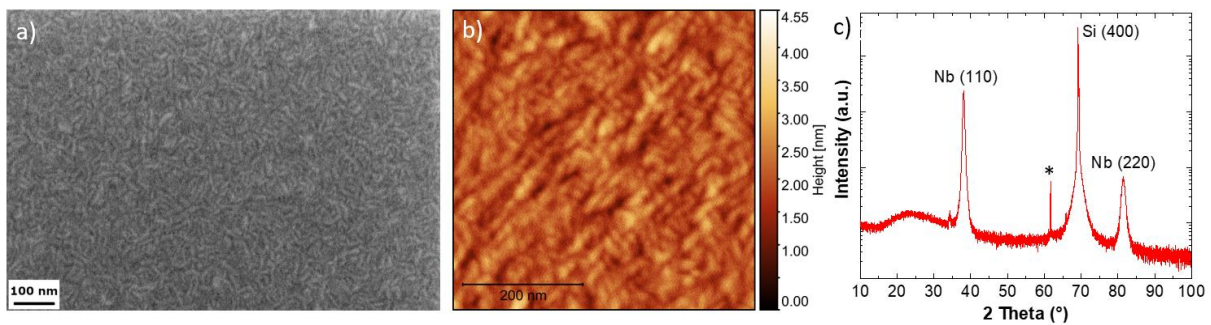

**Figure 5: a) SEM micrograph of the Ti/Nb thin film. b) AFM morphological analysis of the Ti/Nb thin film. c) XRD pattern of the Ti/Nb thin film. The peak highlighted with an asterisk is the $K_β$ component of the intense Si (400) peak.**

An accurate morphological and structural investigation of the Ti/Nb films has been carried out in order to have additional insights for understanding the visible light transport. Fig 5a shows the morphology of the Ti/Nb thin film obtained by scanning electron microscopy (SEM). The film is composed of elongated grains. This peculiar morphology has been widely reported for niobium films obtained by sputtering [45,46]. The presence of a minority percentage (less than 3%) of round grains in the Ti/Nb film is observed. A high magnification SEM micrograph with evaluation of the grain size is reported in Figure S5 of the supporting information. In addition, a comparison of the morphology of the Ti/Nb and Cr/Au morphology is reported in Fig. S6, highlighting the difference in grain size and shape. A 60° tilted SEM is reported in Figure S7 for the evaluation of the quality of the Nb stripe sidewalls. The AFM analysis (Fig. 5b) presents the elongated grains morphology, previously shown in the SEM image. The root mean square (RMS) roughness is evaluated as 460 pm.

X-ray diffraction (XRD) allows identifying a sizeable preferential orientation of the grains composing the film and provides an estimate of the grain size by well-established methodology. The Ti/Nb thin film (Fig. 5b) presents a (110) preferential orientation of the grains, demonstrated by the appearance of the sole (110) and (220) peaks at about 38° and 82°, respectively. Considering the nanocrystallinity of the film, the average grain size is evaluated by the fundamental parameter method [47], giving an estimate of about 25 nm for the Ti/Nb film. A comparison of the XRD patterns with the standard peaks of niobium is reported in Fig. S8.

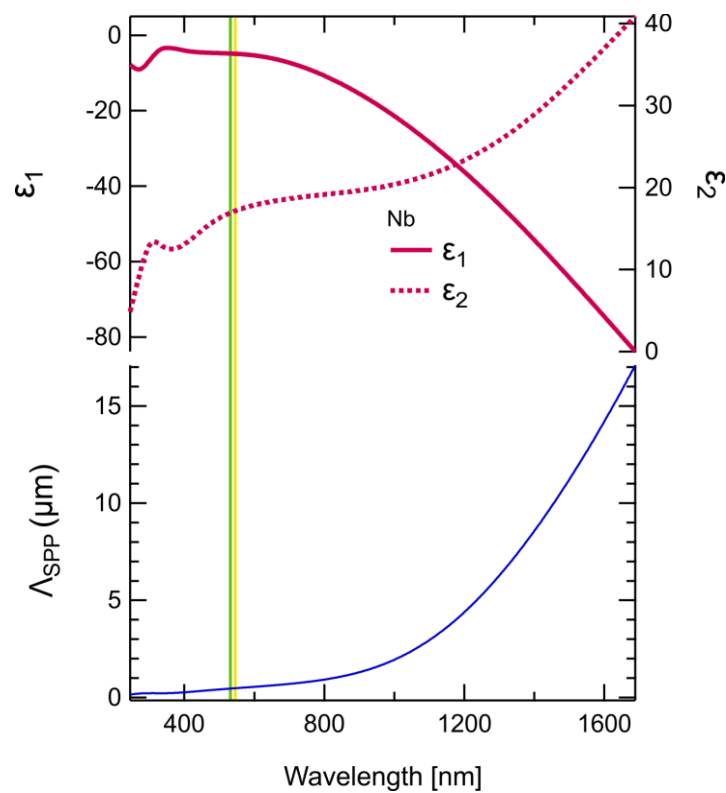

**Figure 6: Top panel, dielectric functions of a Nb film. Bottom panel, attenuation length of SPP at the interface between Nb and PMMA. The vertical green line is at λ=532 nm (excitation), while the yellow line is at λ=546 nm (Raman).**

In Fig. 6, top panel, we report the dielectric functions of Nb film, extracted from spectroscopic ellissometry (SE) measurements on a thick, reference Nb film (purple lines). The dielectric functions are representative of a lossy metal, as testified by the relatively high values of $\varepsilon_2$ even in the visible range. In Fig. 6, bottom panel, we report the attenuation length of SPP at the interface between Nb and PMMA, calculated according to the model proposed by Derrien et al [48]. Around 546 nm, the attenuation length reads around 480 nm. The theoretical value for the attenuation length is perfectly compatible with the experimental observations (600 nm), thereby supporting the assignment of the light transport to the excitation of SPP in the Ti/Nb stripes. We notice that calculated SPP attenuation length for Au at 546 nm read 300 nm, again in perfect agreement with the data for Cr/Au stripes.[48] The theoretical evaluation of the skin depth for the different adhesion layer is reported in Fig S9, showing that the intensity of the incident electromagnetic field completely decays within few tens of nm from the interface.

**Discussion**

The possible mechanisms of the light transport along the stripes are multiple. The light might propagate in the polymer layer due to a waveguide effect, with the coupling into the objective aided by scattering and reflections at the metal surface, or it can propagate along the stripes via the excitation of SPP, either at the bottom $SiO_2$/Nb interface or the top Nb/PMMA interface. We have ruled out the first hypothesis, carrying out similar experiments on gold stripes, which revealed no waveguiding effect of the bP Raman signal. In case of the light transport via SPP, despite being a lossy material, Nb has indeed recently been exploited for SPP-mediated light propagation [20]. Under appropriate circumstances, it actually exhibits some advantages with respect to more-common materials like Ag or Au (e.g. higher adhesion to glass substrates). Niobium is typically employed as plasmonic material in the near and short-wavelength infrared. In fact, Nb plasmonic resonances can range from 1500 nm for thick films [21] up to 3 μm for engineered nanoantennas [22]. We stress that, in our experimental geometry, the SPP excitation is performed at the top interface (PMMA/Nb) but the SPP propagation involves all the stripe, since a finite value of the field is needed at the (bottom) Nb/bP interface.

An additional aspect that is worth mentioning is the difference between the experimental and theoretical values of the attenuation length. The theoretical value of the attenuation length at 545 nm (Raman line) obtained using the model of Derrien et al. [48] , is 480 nm, while the experimental value obtained in our experiments is 600 nm. The enhanced attenuation length can be played by the nanocrystallinity and surface roughness of the Nb thin film, with the particular elongated morphology of the grains. In fact, previous work on SPP on gold surfaces with nanometric roughness has demonstrated an enhancement of the attenuation length [49].

**Conclusions**

In conclusion, we demonstrated the transport of visible light in nanometric and nanocrystalline Nb stripes coupled to a dielectric polymeric layer, using the remote excitation and remote detection of the Raman signal of a black phosphorus flake. The generation of surface plasmon polaritons at the interface between the metal and the dielectric is identified as the dominant mechanism of the light propagation. The detection length is limited to 2.6 μm by the lossy nature of the niobium in the visible range. The accurate characterization of the morphological and structural properties of the niobium film reveals a nanostructured film composed of elongated grains. The optical properties of reference Nb films reveal that their optical behavior is compliant with literature reports, yet that SPP propagation is still achievable. These findings open new perspectives in the use of Nb as an optically active layer, when coupled to an overlying polymer. This technique, which has been applied to bP in this work, could be extended to graphene and other materials whose coupling with Nb is relevant for quantum technologies[50,51].

**Acknowledgements**

This work has benefited from the equipment and framework of the COMP-HUB Initiative, funded by the 'Departments of Excellence' program of the Italian Ministry for Education, University and Research (MIUR, 2018-2022). Dr. A. Camposeo is acknowledged for useful discussions.

**Supporting Information**

**Propagation of Visible Light in Nanostructured Niobium Stripes coupled to a dielectric polymer**

F. Telesio[1], F. Mezzadri[2,3], M. Serrano-Ruiz[4], M. Peruzzini[4], F. Bisio[5], S. Heun[1], and F. Fabbri[1]

1 NEST, Istituto Nanoscienze – CNR and Scuola Normale Superiore, Piazza San Silvestro 12, 56127 Pisa, Italy

2 IMEM-CNR, Parco Area delle Scienze 37/a, Parma, 43124, Italy

3 Department of Chemistry, Life Sciences and Environmental Sustainability, University of Parma, Parco Area delle Scienze 11/A, 43124 Parma, Italy

4 CNR-ICCOM, Via Madonna del Piano 10, 50019 Sesto Fiorentino, Italy

5 CNR-SPIN, C.so Perrone 24, 16152 Genova, Italy

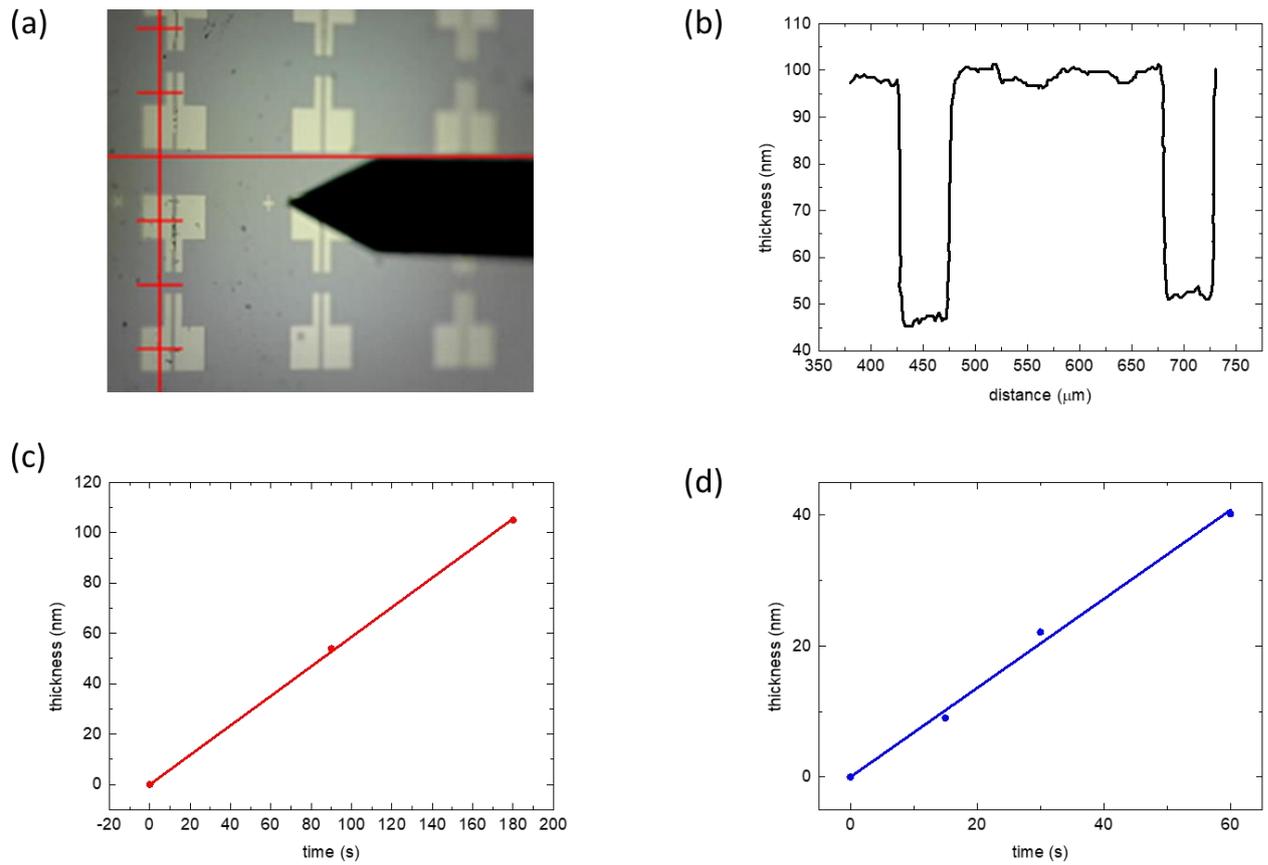

**Figure S1: Profilometry analysis and thickness evaluation of the Nb film. (a)** Optical image of a calibration sample with an optical lithography pattern with lateral size of several hundreds of μm. In the image, the shadow of the optical profilometer tip is visible. **(b)** Thickness profile of calibration sample obtained from a stylus profilometer measurement. **(c)** Calibration of the Nb sputtering rate. The obtained rate is 0.6 nm/s. **(d)** Calibration of the Ti sputtering rate. The obtained rate is 0.7 nm/s. Each calibration point is the average of the film thickness evaluation on several profiles. The sputtering parameters are 4 mTorr of Ar and approximately 150 W.

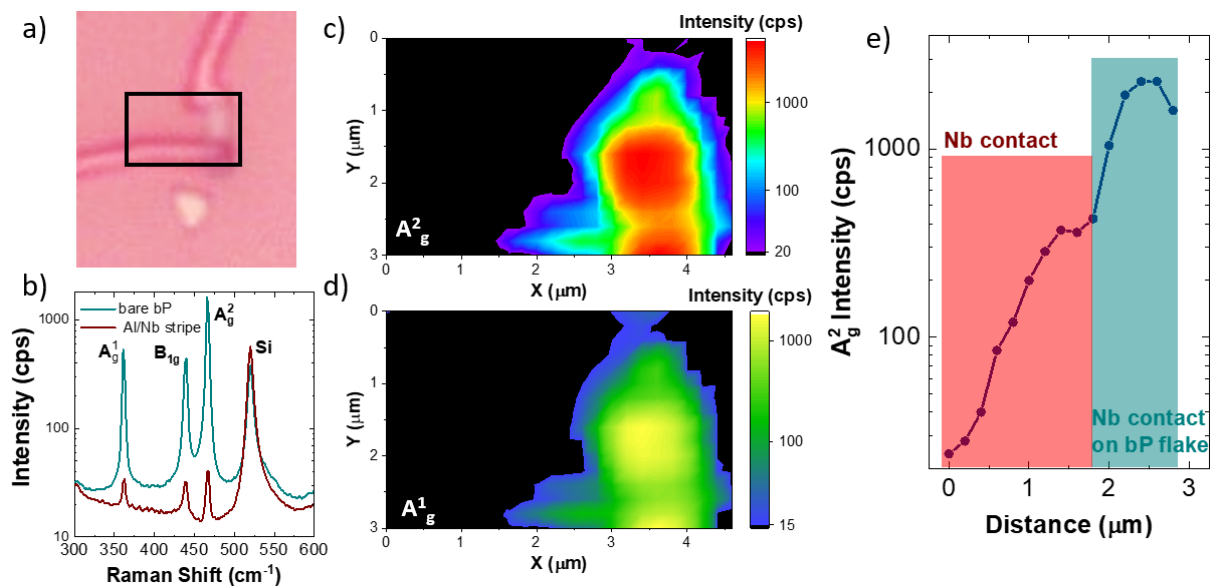

**Figure S2:** a) Optical microscopy image of a bP flake with Al/Nb stripes. The black square indicates the area where the Raman maps were acquired. b) Raman spectra acquired on the bare bP (red line) and on the Al/Nb stripe 1 μm away from the bP flake (blue line). c) Intensity map of the A2g mode. d) Intensity map of the A1g mode. e) Line profile of the $A^2_g$ intensity along the Al/Nb stripe in logarithmic scale.

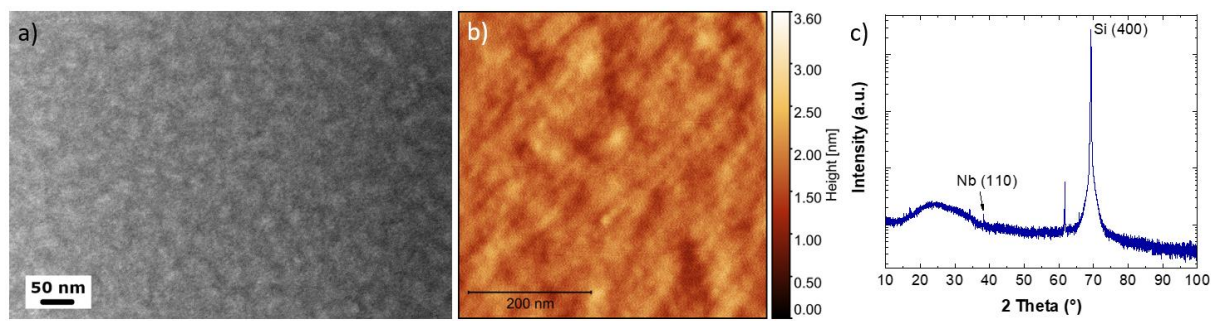

**Figure S3:** (a) SEM micrograph, (b) AFM morphological analysis. and (c) XRD pattern of an Al/Nb thin film.

The SEM imaging and the XRD analysis reveal that the Al/Nb film is poorly crystalline. The Al/Nb film presents a RMS roughness of 370 pm.

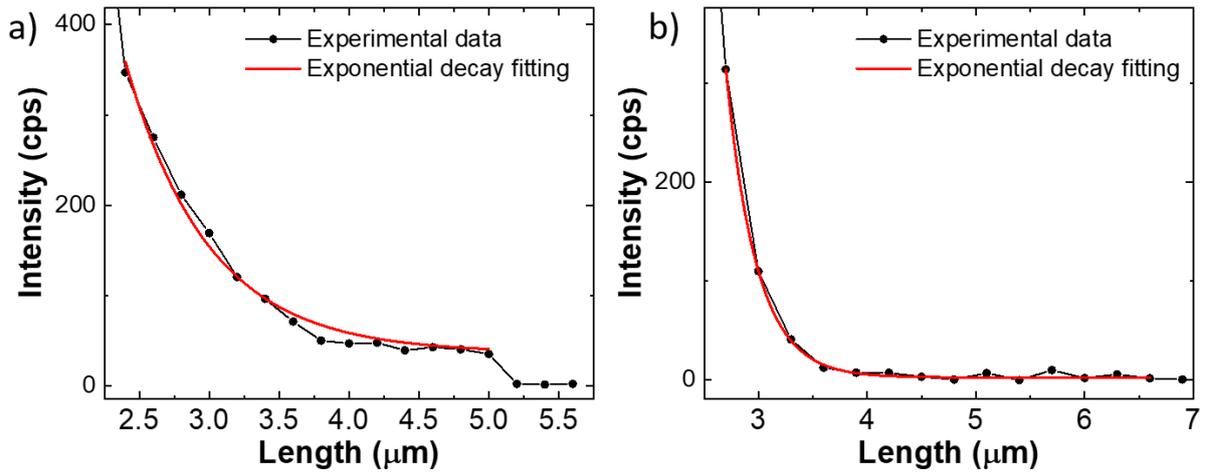

**Figure S4:** exponential decay fitting of the data presented in Fig. 4, in case of Ti/Nb (a) and Cr/Au (b) for the evaluation of the propagation length.

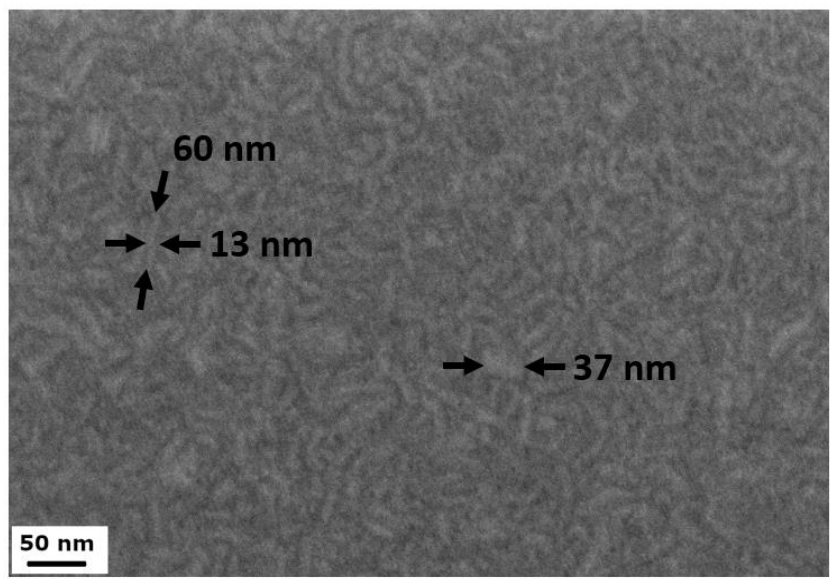

**Figure S5:** High magnification SEM micrograph of a Ti/Nb thin film. The high magnification SEM image allows a detailed evaluation of the grain size. Here, the elongated grains have a size of 60 nm x 13 nm, while the round grains have a diameter of 37 nm.

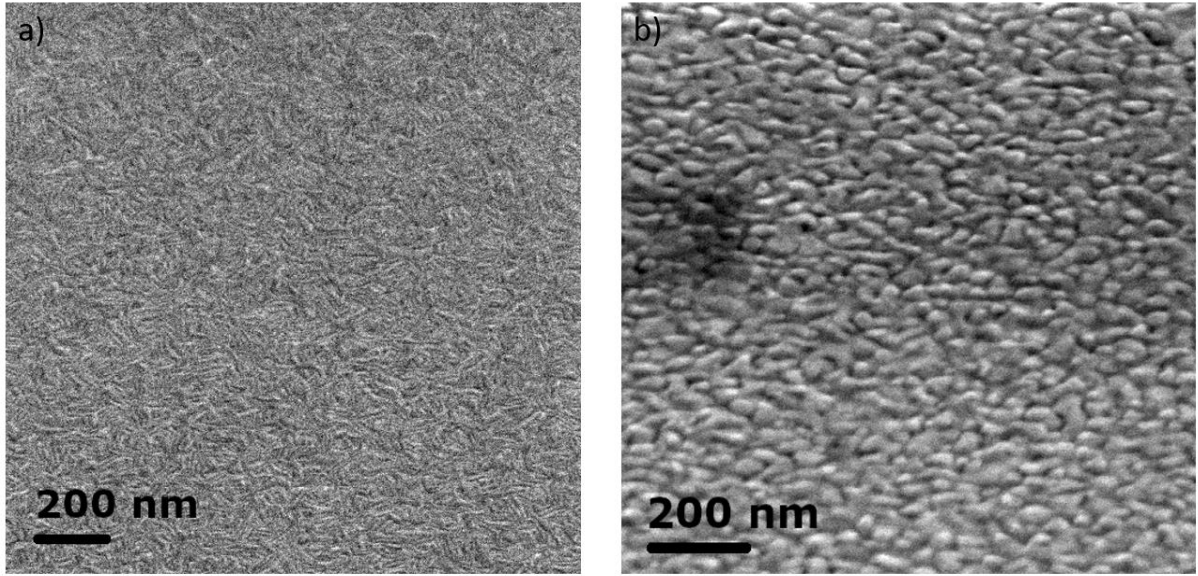

**Figure S6:** Tilted high magnification SEM micrographs of a Ti/Nb (panel a) and Cr/Au (panel b) thin films. The high magnification SEM images allow the comparison of the morphology of the two different thin films. The Ti/Nb thin film morphology shows the elongated grain morphology, previously reported while in case of the Cr/Au, the film presents a morphology composed of round grains with a size varying between 40 nm and 60 nm.

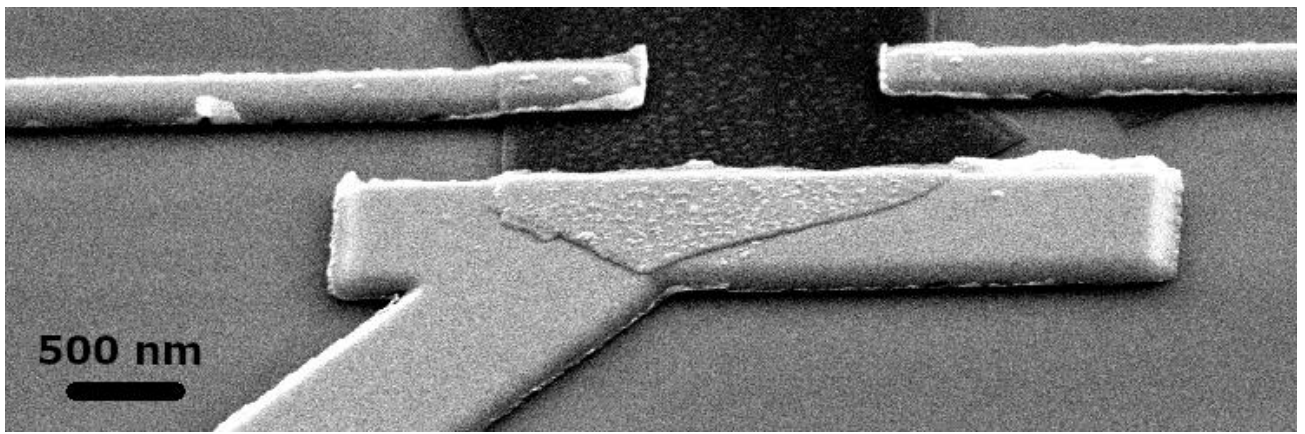

**Fig. S7:** 60° tilted SEM image of the Ti/Nb contact after the removal of the PMMA layer. The PMMA was removed by a wet process using warm acetone and a gentle rinsing in isopropanol alcohol. The SEM analysis shows that the sidewall of the Nb stripes presents tilted sidewalls. The Nb thickness decreases over a 100-130 nm length, therefore the sidewall angle with respect to the substrate perpendicular is between 28° and 35°.

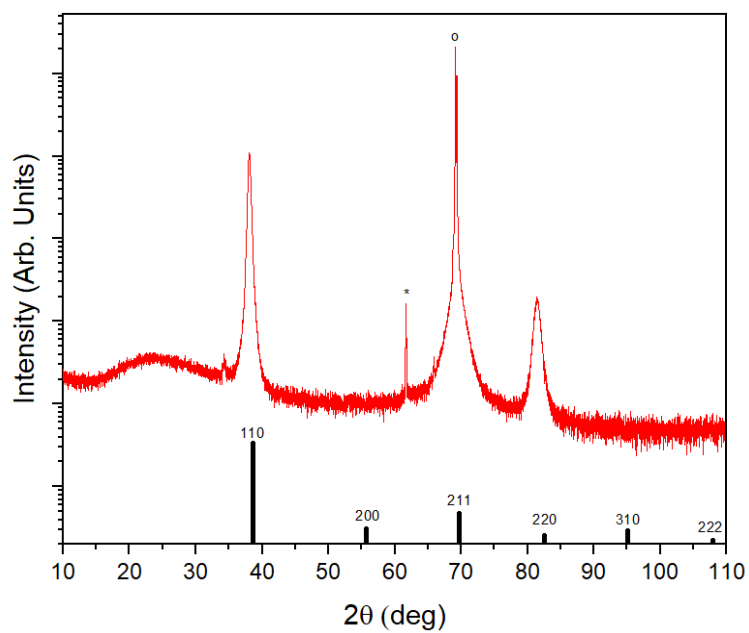

**Figure S8: comparison of the PXRD pattern of the Ti/Nb thin film (Log scale) with the tabulated one for niobium (ICDD # 00-034-0370, linear scale). The "o" symbol indicates the silicon substrate (400) reflection., The asterisk corresponds to the minor K$_\beta$ component of the Si (400) peak, not avoidable for intense reflections in the present experimental conditions.**

It is worth noting that the cell parameter is slightly different from the tabulated value ($a$ =3.3066 Å). In fact, in case of the Ti/Nb film, the cell parameter is $a$ =3.34358(11) Å. These slightly larger values can be due to strain effects and the nanocrystallinity of the film.

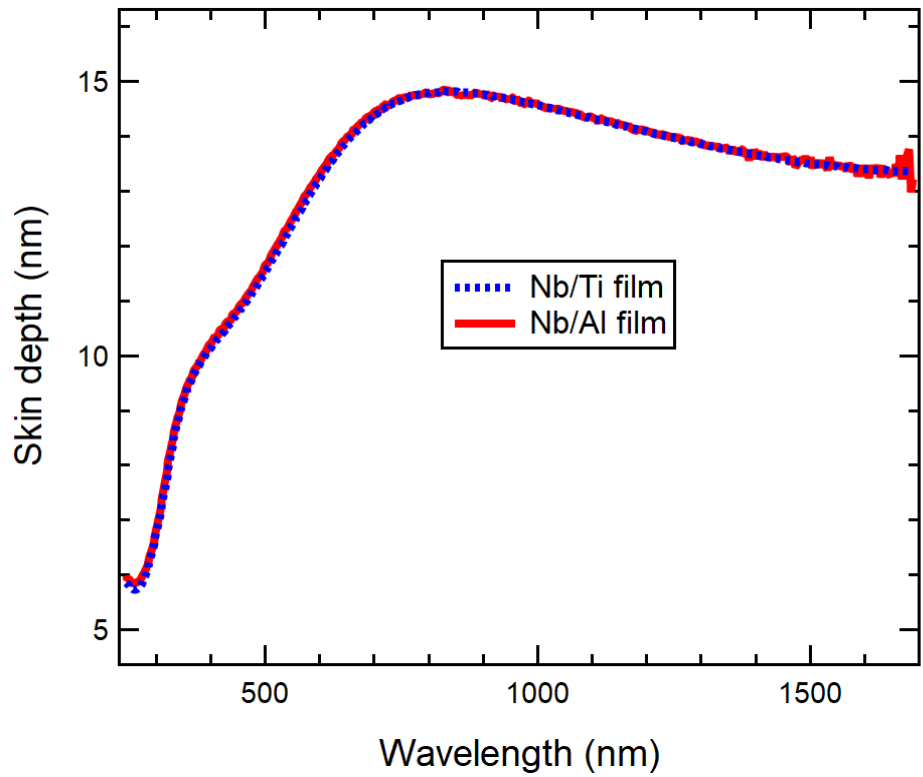

**Figure S9:** Skin depth of the light propagation in the Ti/Nb and Al/Nb films derived from the measured dielectric constant. The skin depth is 12 nm for both the adhesion layer at the 532 nm and 546 wavelengths.